\documentclass[journal,final]{IEEEtran}
\usepackage[pdftex]{graphicx}
\usepackage{epstopdf}
\DeclareGraphicsExtensions{.pdf,.jpeg,.png,.eps}
\usepackage{cite}
\usepackage{amssymb}
\usepackage{amsmath}
\usepackage{graphics}
\usepackage{graphicx}
\usepackage{float}
\usepackage[T1]{fontenc}
\usepackage[utf8]{inputenc}
\usepackage{authblk}
\usepackage{authblk}
\usepackage{tabularx}
\usepackage{balance}
\usepackage{ amssymb }
\usepackage{algorithm2e}
\usepackage{algorithmic}
\usepackage{gensymb}
\usepackage{subfig}
\usepackage{multicol}
\usepackage{lipsum}
\usepackage{chngpage}
\usepackage{calc}
\usepackage{ragged2e}
\usepackage{tabularx}
\begin{document}
%\markboth{Submitted to IEEE Journal of Lightwave Technology }{Marshoud \MakeLowercase{\textit{et al.}}: Optical Adaptive Precoding for   Visible Light Communications}
\title{Optical Adaptive Precoding for   Visible \\ Light Communications}

\author{
Hanaa~Marshoud,~\IEEEmembership{Student~Member,~IEEE}, Paschalis C. Sofotasios,~\IEEEmembership{Senior~Member,~IEEE}, Sami~Muhaidat,~\IEEEmembership{Senior~Member,~IEEE},~Bayan~S.~Sharif,~\IEEEmembership{Senior~Member,~IEEE},~and~George~K.~Karagiannidis,~\IEEEmembership{Fellow,~IEEE}

%\thanks{Manuscript received xxx; revised xxx.}%
\thanks{H.~Marshoud  and B. S. Sharif  are with the Department of Electrical and Computer Engineering, Khalifa University, PO Box 127788, Abu Dhabi,  UAE (email:  $\{\rm hanaa.marshoud; bayan.sharif \}@kustar.ac.ae$).}%
\thanks{P. C. Sofotasios is with the Department of Electrical and Computer Engineering, Khalifa
University, PO Box 127788, Abu Dhabi, UAE, and  with the Department of Electronics and Communications Engineering, Tampere University of Technology, Tampere FI-33101, Finland (e-mail: $ \rm p.sofotasios@ieee.org$).}%
\thanks{S.~Muhaidat is with the Department of Electrical and Computer Engineering, Khalifa University, PO Box 127788, Abu Dhabi, UAE   and with the Institute for Communication Systems, University of Surrey, GU2 7XH, Guildford, UK  (email: $\rm muhaidat@ieee.org$).}%
\thanks{G.~K.~Karagiannidis is with  the Department of Electrical and Computer Engineering, Aristotle University of Thessaloniki, 54 124 Thessaloniki, Greece (email:  $ \rm geokarag@auth.gr$).}
%\thanks{Copyright (c) 2015 IEEE. Personal use of this material is permitted. However, permission to use this material for any other purposes must be obtained from the IEEE by sending a request to pubs-permissions@ieee.org.}%
%\thanks{Digital Object Identifier $\#\#\#$}%
}

\maketitle

\begin{abstract}
\par %One of the key challenges in visible light communication (VLC) is the limited modulation bandwidth of the light sources, which typically spans a few MHz. Recently,
Multiple-input multiple-output (MIMO) techniques  have  recently demonstrated significant   potentials   in  visible light communications (VLC),  as they can overcome the modulation bandwidth  limitation and provide  substantial improvement in  terms of  spectral efficiency and link reliability.
However, MIMO systems typically suffer from inter-channel interference, which causes severe degradation to the  system performance. In this context,  we propose a novel optical adaptive precoding (OAP) scheme
for  the downlink   of MIMO VLC systems, which exploits the knowledge of  transmitted symbols  to enhance the  effective signal-to-interference-plus-noise ratio. We also  derive    bit-error-rate expressions for the OAP under perfect and outdated channel state information (CSI). Our results  demonstrate   that the proposed scheme is  more robust to both CSI error and channel correlation, compared to  conventional channel inversion precoding.
\end{abstract}
\vspace{-0.02cm}

\vspace{-0.17cm}

\section{Introduction}

\label{sec:model}
\begin{figure*}[\widetilde{}]
\vspace{-2cm}
\label{fig:system_model}
\center\includegraphics[width=6.5in,height=3.8in]{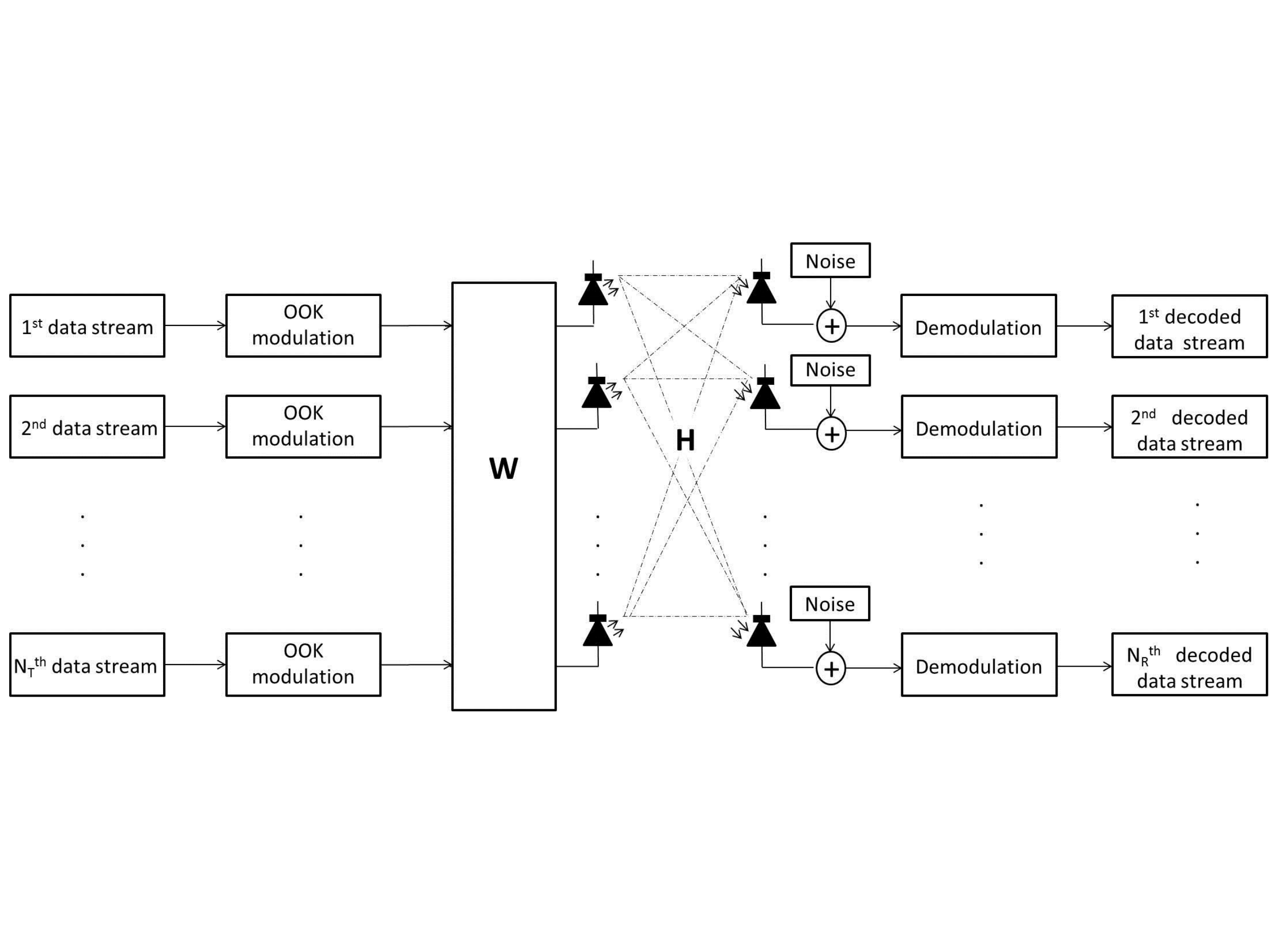}
\vspace{-2cm}
\caption{ MIMO-VLC system.}
\label{fig:system_model}
\end{figure*}
With the explosive  growth of broadband applications and Internet services, the consumer demands for  seamless mobile data connectivity  increase rapidly.
Given that  the majority of wireless data traffic is generated indoors \cite{cisco_indoor}, service providers are constantly looking for  innovative  solutions to provide  robust indoor wireless coverage. In this context, the concept of small cells and heterogeneous networks has been  proposed as a potential solution to this problem. Nevertheless, the imminent shortage of radio frequency (RF) resources at large, along with  the  evidenced spectral
efficiency saturation, has revealed  the need for larger bandwidth and spectral relief.

To address some of the aforementioned challenges,    visible light communication (VLC) has recently  emerged as a promising solution to support and complement traditional RF communication systems, owing to  its capability to  overcome the currently witnessed scarcity of  the radio spectrum resources. To this effect, VLC is opening a new pathway for connecting indoor users to the Internet by practically  unlocking the untapped electromagnetic spectrum in the visible light region \cite{VLC_mag}. The interest in this technology is driven by several factors as it offers high degree of spatial reuse, enhanced communication security and increased energy efficiency. Moreover, unlike RF signals, VLC does not interfere with electromagnetic radiations, which makes it safe to be used in places with  high electromagnetic interference (EMI), such as hospitals and industrial plants \cite{market,survey2}.

VLC uses light emitting diodes (LEDs) as transmitters and photo detectors (PDs) as receivers. Due to the relatively high modulation bandwidth of the LEDs, the amplitude
of the optical signals can be modulated at sufficiently high switching rate that cannot be perceived by the human eye. Thus, the illumination function of the LEDs is not affected. This process is known as intensity modulation (IM), and it is feasible due to the incoherent light emitted by the LEDs, where photons have different wavelengths and phases, unlike coherent light sources such as lasers. At the receiver site, direct detection (DD) of the arriving signals is performed to convert the received optical signal into electric current. It is recalled that direct detection is substantially simpler  than coherent detection used in RF, where an oscillator is required to extract the baseband signal from the carrier. Also, the diameters of the PDs are typically several hundred times the wavelength of light signals and thus, spatial multi-path interference is mitigated by the integrating effect at the PDs' surfaces \cite{haas_book,murat2,state-of-the-art}.

It has been recently shown that multiple-input-multiple-output (MIMO) systems can contribute towards  enhancing  the spectral efficiency of indoor VLC systems \cite{MIMO1,MIMO2,MIMO5,MIMO7,MIMO8,MIMO_VLC_Haas}. %Motivated by the fact that indoor spaces are typically equipped by multiple  light emitting diodes (LEDs), several optical MIMO techniques have been devised to facilitate high data rate transmission over VLC channels \cite{MIMO_VLC_Haas}. %Furthermore, it has been  demonstrated that  MIMO configurations can reduce the difficulties in achieving physical alignment between the transmitting LEDs and the receiving photo detectors (PDs) in scenarios where the receivers move around the coverage area of the LEDs \cite{MIMO1}.
As demonstrated in the open literature, the  implementation of MIMO schemes  in  multi-user (MU) scenarios  creates  MU interference, which has a detrimental  impact on the overall system performance.  Hence, interference mitigation techniques  are essential and  transmit precoding  techniques constitute an effective method, which is widely considered  in downlink transmissions \cite{marshoud}.

Various transmit precoding schemes have been recently proposed aiming to eliminate the incurred inter-channel interference (ICI) in indoor MU-MIMO VLC systems  \cite{BD_VLC,THP}. Motivated by this,  in the present contribution we propose a design framework  for a novel and effective  optical adaptive precoding (OAP) scheme. Specifically, unlike most existing approaches, where precoding is employed  to completely eliminate interference, the proposed  precoder  exploits   the knowledge of  constructive interference  between spatial streams in order to enhance the overall system performance. This scheme is based on on-off keying (OOK) modulation, which is widely considered  in VLC systems \cite{ook_standard}. Note that a  similar concept was considered in the context of radio frequency (RF) MIMO systems  \cite{Dynamic1,Dynamic2}, where it led to performance gains compared to conventional channel inversion (CI) precoding.
%%Why
It is evident that OAP is particularly suitable in  the context of VLC systems for the following reasons:
\begin{itemize}
\item The channel coefficients in indoor VLC remain constant, unless a change in the user position occurs. Thus, given that the indoor mobility velocity is relatively  low, we can assume that the precoder calculation can be performed once per symbol, which leads to   lower complexity overhead, compared to conventional  RF systems.
\item Optical channels are real and positive, which allows the estimation of  the received interference,    based only on the transmitted symbols. Hence, adaptive precoding in VLC becomes robust to channel state information (CSI) errors.
\item Given that  ICI is not completely eliminated, the high correlation between the optical subchannels  is  practically beneficial, when the transmitted symbols are equal.
\end{itemize}

To the best of our knowledge, this is the first work that proposes the exploitation of constructive interference to enhance  the performance of VLC systems. In addition, it  quantifies the impact of outdated CSI  that results from the mobility of indoor users in the proximity of the  LEDs, leading to erroneous generation of the precoding matrix at the transmitter and, thus, to  inaccurate detection at the receiver  \cite{coordinated}. Specifically, the contribution of the present paper is summarized below:
\begin{itemize}
\item We propose an optical adaptive precoding scheme that utilizes  the knowledge of the transmitted symbols  to constructively correlate the interference  in an indoor MU-MIMO VLC downlink network.
\item We investigate the bit-error-rate (BER)  performance of the proposed scheme assuming  perfect and outdated CSI, where we derive a closed-form expression for the former and an upper bound for the latter.
\item We provide  extensive simulation results to demonstrate the validity of the derived analytic expressions and extract useful  insights  on  the performance of the proposed scheme under different realistic scenarios.
\end{itemize}

It is shown  that the  proposed OAP scheme introduces significant   performance improvement   compared to conventional channel inversion, while the corresponding  added complexity is minimal.  Furthermore, it  is particularly  robust to outdated CSI error, which result due to the mobility of users, and to the correlation between the channels of   different users.

The remainder of the paper is organized as follows: Section \ref{sec:model} introduces the system and channel model of MU-MIMO VLC. Section \ref{sec:precoding} presents the  conventional channel inversion  and the proposed OAP, along with the  corresponding BER analysis for both schemes under perfect and outdated CSI. The respective numerical  results are presented in Section \ref{sec:results}, while closing remarks are provided in Section \ref{sec:conc}.

%\emph{Notation:} $(.)^{T} ,(.)^{*} and (.)_{ik}$ denote transpose,  Hermitian transpose operations and the $(i,k)^{th}$ entry of a matrix, respectively.

\section{System and  Channel Model}
\label{sec:model}

\subsection{System Model }

We consider a  generic VLC system with $N_T$ transmitting LEDs and $N_R$ receiving  PDs, as illustrated in F\mbox{}ig. \ref{fig:system_model}, where  the employed PDs may correspond to single or multi users with no impact on the presented analysis. Based on this,
%OOK modulation is applied due to its popularity in VLC systems \cite{ook_standard}.
the received signal can be represented as
\begin{equation}
\textbf{y}= \textbf{H}\textbf{x}+ \textbf{n}
\end{equation}
where $\textbf{H} \in \mathbb{R} ^{N_R \times N_T}$ is the MIMO channel, namely
%\begin{equation}
%\textbf{H}= \begin{bmatrix}
%\underbrace{h_{11}}_{desired} & \underbrace{h_{12}}_{interference} & \ldots & \underbrace{h_{1N_T}}_{interference}\\
%\underbrace{h_{21}}_{interference} & \underbrace{h_{22}}_{desired} & ... & \underbrace{h_{2 N_T}}_{interference}\\
%\vdots & \vdots & \vdots & \vdots\\
%\underbrace{h_{N_R 1}}_{interference} & \underbrace{h_{N_R 2}}_{interference} & \ldots& \underbrace{h_{N_R N_T}}_{desired} \\
%\end{bmatrix}.
%\end{equation}
\begin{equation}
\textbf{H}= \begin{bmatrix}
{h_{11}}_{D} & {h_{12}}_{I} & \ldots & {h_{1N_T}}_{I}\\
{h_{21}}_{I} & {h_{22}}_{D} & ... & {h_{2 N_T}}_{I}\\
\vdots & \vdots & \vdots & \vdots\\
{h_{N_R 1}}_{I} & {h_{N_R 2}}_{I} & \ldots& {h_{N_R N_T}}_{D} \\
\end{bmatrix}.
\end{equation}
where ${h_{ij}}_D$ and ${h_{ij}}_I$ indicate the desired and interference channel paths, respectively. Hence, the received signal at the $i^{th}$ PD can be expressed as
\begin{equation}
y_{i}=\gamma P \textbf{h}^{T}_{i} \textbf{x} +\textbf{n}_i
\end{equation}
where  $\gamma$ is the detector responsivity,  $P$ is the LED transmitting power, $\textbf{h}^{T}_i$ denotes the  $i^{th}$ row in  $\textbf{H}$ and $\textbf{x}$ is the transmitted signal. Also, $\textbf{n}_i$  is the noise vector with statistically independent and identically distributed  entries drawn from a circularly-symmetric Gaussian distribution of zero mean and variance,
\begin{equation}
\sigma_i^2 = \sigma_{sh}^2 + \sigma_{th}^2
\end{equation}
with $\sigma_{sh}^2 $ and $ \sigma_{th}^2$ denoting  the variances of the shot noise and the thermal noise, respectively.
The shot noise in an optical wireless channel results from the high rate physical photo electronic conversion process,  with variance at the $i^{th}$ PD
\begin{equation}\label{equ:shot}
\sigma_{sh_i}^2=2qB\left(\gamma\sum\limits_{j=1}^{N_T}h_{ij}x_{j}+I_{bg}I_2 \right)
\end{equation}
\noindent where \textit{q} is the electronic charge, \textit{B} is the bandwidth, $I_{bg}$ is background current, and $I_2$ is the noise bandwidth factor. Moreover, the thermal noise is generated within the transimpedance receiver circuitry, and its variance can be determined by
\begin{equation}\label{equ:thermal}
\sigma_{th_i}^2 = \dfrac{8 \pi k T_k}{G}\eta AI_2B^2 + \dfrac{16 \pi^2kT_k \Gamma }{g_m}\eta^2 A^2 I_3 B^3.
\end{equation}
\noindent In (\ref{equ:thermal}), \textit{k} is Boltzmann's constant, $T_k$ is the absolute temperature, \textit{G} is the open-loop voltage gain, \textit{A} is the PD area, $\eta$ is the f\mbox{}ixed capacitance of the PD per unit area, $\Gamma$ is the field-effect transistor (FET) channel noise factor, $g_m$ is the FET transconductance, and $I_3$ = 0.0868 \cite{fundamental}.

Based on the above, the corresponding signal-to-interference-plus-noise ratio (SINR) at the $i^{th}$ PD is expressed as
\begin{equation}
\textrm{SINR}_{i}=\frac{\gamma  P h_{ii}}{\gamma  P \sum_{j=1, j\neq i}^{N_T} h_{ij} + {2\sigma_i}}.
\end{equation}

\subsection{Channel Model}
A line of sight (LOS)  VLC channel is naturally assumed  and the  coefficients of the channel matrix $\textbf{H}$ are given by
\begin{equation}\label{equ:hij}
h_{ij}= \begin{cases}\frac{A_i}{d^2_{ij}}R_o(\varphi_{ij})T_s(\phi_{ij})g(\phi_{ij})\cos(\phi_{ij})& \quad  0\leq\phi_{ij}\leq \phi_{c}\\
0& \quad  \phi_{ij}>\phi_{c}
\end{cases}
\end{equation}
\noindent where $i =1, 2, 3, \ldots  , N_R$, $j =1, 2, 3, \ldots  , N_T$,   $A_i$ denotes the receiver PD area, $d_{ij}$ is  the distance between $j^{th}$ transmitting LED  and  $i^{th}$ receiving PD, $\varphi_{ij}$ is the angle of emergence with respect to the transmitter axis, $\phi_{ij}$  is the angle of incidence with respect to the receiver axis, $\phi_{c}$ is the field of view (FOV) of the PD, $T_s(\phi_{ij})$ is the gain of optical filter and $g(\phi_{ij})$ is the gain of the optical concentrator, which is expressed as

\begin{equation}\label{equ:g}
g(\phi_{ij}) = \begin{cases} \frac{n^2}{\sin^2(\phi_{c})} \qquad \qquad & 0\leq\phi_{ij}\leq \phi_{c}\\
 0 \qquad &\phi_{ij}>\phi_{c}
 \end{cases}
\end{equation}
\noindent where $n$ denotes the corresponding  refractive index.  Moreover,
\begin{equation}\label{equ:R_o}
R_o(\varphi_{ij}) = \frac{m+1}{2\pi}\cos^m(\varphi_{ij})
\end{equation}
denotes the Lambertian radiant intensity of the transmitting LEDs, where
 \begin{equation}
 m = \frac{-  \ln(2)}{\ln(\cos(\varphi_{1/2})}
 \end{equation}
 % is  the order of Lambertian emission,
%\begin{equation}\label{equ:m}
% m =\frac{\ln(2)}{\ln(\cos(\varphi_{1/2}))}
%\end{equation}
with $\varphi_{1/2}$ representing  the corresponding  transmitter semi-angle at half power.

\subsection{Outdated CSI}
While the channel in VLC is considered deterministic for given transmitter-receiver specifications and fixed locations, the assumption of perfect CSI at the transmitters is not in fact realistic even for indoor VLC systems. This is because the  mobility of users in indoor environments, such as  museums, libraries or large offices, can change the channel coefficients of the users. Thus, it is evident that if this change occurs between CSI updates, it results to outdated CSI error.  %Typically, CSI at the receivers can be obtained by means of periodic pilot signals, then the receivers feed back the quantized channel coefficients to the transmitters by an RF of IR uplink\footnote{Although VLC uplink is also possible, it is energy-inefficient for low-power mobile devices. Thus, utilizing uplink-downlink reciprocity for acquiring  CSI at the transmitter is not relevant for VLC.}.  The uncertainty in the VLC channel estimation arises from the noise in the downlink and uplink channels as well as the mobility of users.
Based on this, in  order to quantify and develop insights on the effect of imperfect CSI on system performance, we assume additive stochastic error for the CSI at the transmitter, namely
\begin{equation}
  \hat{\textbf{H}}=\textbf{H}+{\epsilon_o}.
  \end{equation}
In the above,  $\hat{\textbf{H}}$ denotes the estimate of $\textbf{H}$ available at the  transmitter and  $\epsilon_{o} \leq \mathcal{E}$, with $\mathcal{E}$ denoting the error bound that occurs when the mobile user moves with maximum velocity between the  reception of pilot signals and data \cite{coordinated}.

\begin{figure}[h]
\center
\includegraphics[width=7cm,height=5cm]{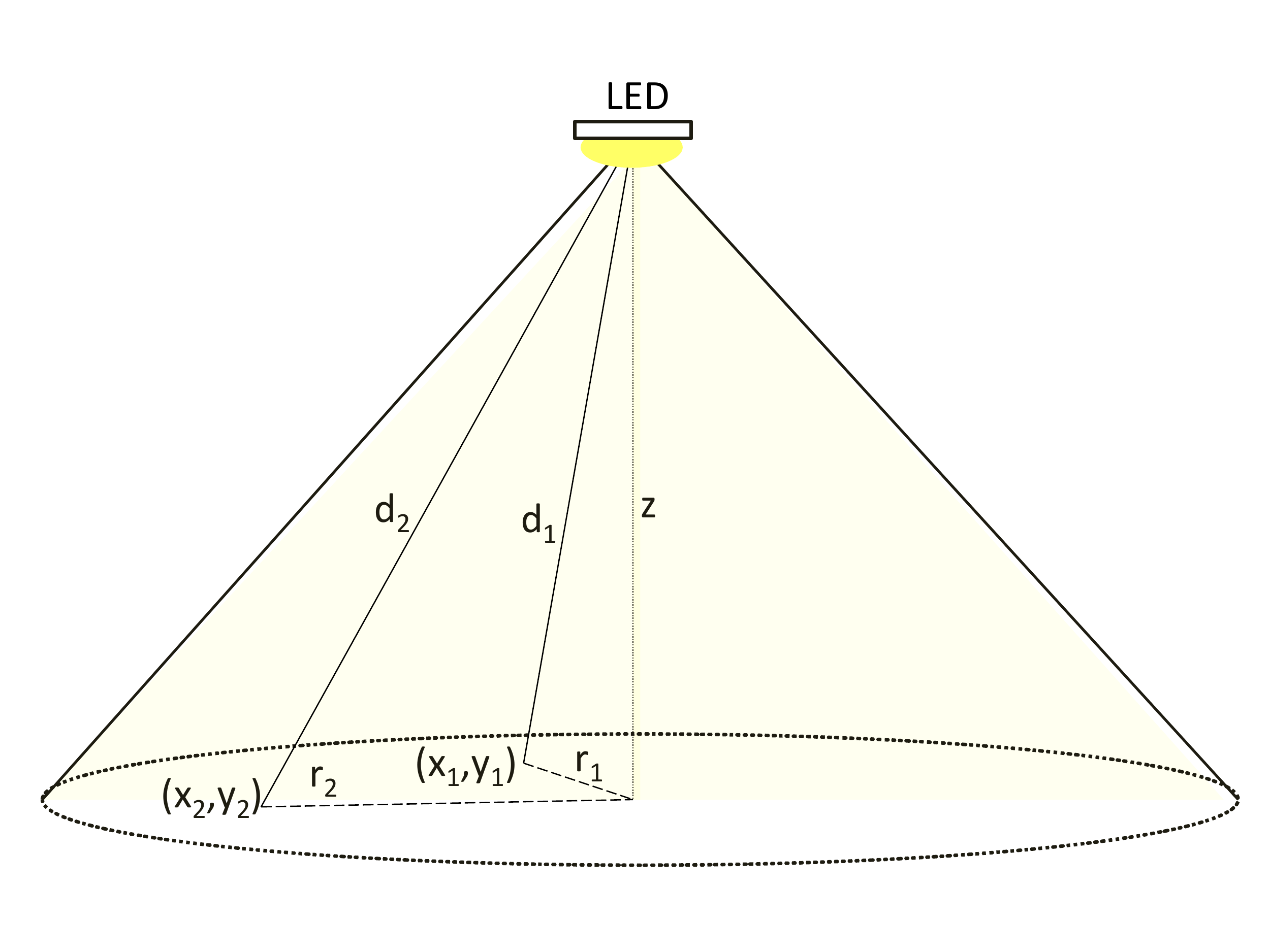}
\caption{Outdated CSI due to user mobility.}
\label{fig:mobility}
\end{figure}

In order to determine the value of $\mathcal{E}$, we simplify the channel gain in (\ref{equ:hij}) using \eqref{equ:g} and \eqref{equ:R_o}. To this end, we substitute  $\cos\varphi_{i}$ with $z/d_{i}$, where $z$ denotes the height between the LEDs' and the PDs' planes, which is assumed to be fixed. To this effect, it follows that the corresponding  channel gain  $h_{ij}$ can be expressed as
\begin{equation}
h_{ij} = \frac{\varpi}{d^{m+3}_{ij}}
\end{equation}
%\begin{equation}\label{equ:channel_simple2}
%h_{ij} = \varpi \frac{1}{d^{m+3}_{ij}}
%\end{equation}
where
\begin{equation}
\varpi=\frac{(m+1) A_i  T_s(\phi_{ij})g(\phi_{ij})}{2\pi}.
\end{equation}
By also assuming that a user $U_i$ moves along the horizontal plane from $(x_1,y_1)$ to $(x_2,y_2)$, with maximum velocity $v$ as illustrated in Fig. \ref{fig:mobility}, the    error bound $\mathcal{E}$ can be determined by
\begin{equation}\label{equ:outdatedCSI3}
\mathcal{E} = \varpi \left| \frac{1}{d^{m+3}_{2}} - \frac{1}{d^{m+3}_{1}} \right|
\end{equation}
where
\begin{equation}
d^2_1 = x^2_1 +  y^2_1 + z^2,
\end{equation}
\begin{equation}
d^2_2 = x^2_2 +  y^2_2 + z^2
\end{equation}
 and
 \begin{equation}
 v = \frac{\sqrt{(x_2-x_1)^2 + (y_2-y_1)^2}}{t}
 \end{equation}
 with $t$ denoting the time elapsed since the last CSI update at the receiver.

\section{Linear Precoding for MU-MIMO VLC}
\label{sec:precoding}
 % The signal vector $X\in \mathbb{R}^{NT\times1}$  is  multiplied by a precoding matrix, $W\in\mathbb{R}^{N_T \times N_T}$  prior to transmission. Let $H \in \mathbb{R}^{N_R \times N_T}$ represent the MIMO channel, the  received signal can be expressed as

It is recalled that the main objective of  transmit precoding is to mitigate the ICI between the different users, in order to establish parallel channels. In the following, we discuss the design of CI precoding in MU-MIMO VLC systems. Then, it is shown that exploiting the knowledge of the ICI results to  benefits from the constructive part of this interference  that ultimately   lead to considerable performance enhancement.
\subsection{Conventional Channel Inversion Precoding}
CI precoding is a relatively simple approach to eliminate ICI in MIMO downlink communications and it  results to  independent subchannels for each receiving terminal. This can be achieved  by multiplying the transmitted signal with a precoder $\textbf{W}$, which is obtained by means of the  generalized inverse of the channel matrix, namely
\begin{equation}
\label {equ:CI_matrix}
\textbf{W}= \beta\textbf{H}^*.(\textbf{H}.\textbf{H}^*)^{-1}
\end{equation}
where the $^*$ symbol denotes Hermitian transpose   and
\begin{equation}
\label {equ:scaling}
\beta= \sqrt{\frac{1}{ \textbf{x}^* .(\textbf{H}.\textbf{H}^*)^{-1} . \textbf{x} }}
\end{equation}
  is a scaling factor that ensures instantaneous normalization of the transmit power \cite{Dynamic1}.
Based on this, the inversion of the  channel crosscorrelation matrix  $\textbf{R}=\textbf{H}.\textbf{H}^*$ in (\ref{equ:CI_matrix})  nulls all the off-diagonal components, which results to   interference free subchannels. Accordingly, the signal-to-noise ratio (SNR) at the $i^{th}$ PD can be expressed as
\begin{equation}
\textrm{SNR}_{i}=\frac{ \gamma  P \textbf{h}^T_i \textbf{w}_i}{ {2 \sigma_i}}
\end{equation}
where  $\textbf{w}_i$ denotes the $i^{th}$ column vector in  $\textbf{W}$.
Nevertheless, this  leads to significant noise enhancement at the receiving PDs, which constitutes the main limitation of CI precoding.
Consequently, the normalized overall achievable  throughput of the MIMO VLC system can be expressed as:
\begin{equation}\label{equ:Th_CI}
\mbox{Th}_{CI}=\sum_{i=1}^{N_R} \log_2 \left(1+\frac{ \gamma  P \textbf{h}^T_i \textbf{w}_i}{ {2 \sigma_i}}\right).
\end{equation}
\subsubsection{CI BER Performance Under Perfect CSI}

The error probability at the $i^{th}$ PD can be expressed as
\begin{equation}\label{equ:BER_CI}
\begin{aligned}
\mbox{Pr}_{e_i}  =&  \frac{1}{2^{N_T}} \sum_{s=1}^{2^{N_T}} \int_{-\infty}^{\frac{1}{2}  \gamma P \textbf{h}^T_i \textbf{w}_{i_s} } \mathcal{N}_{y_i}\left( \gamma P \textbf{h}^T_i \textbf{w}_{i_s} , \sigma^2_i\right) {\rm d}y_i \\& +   \frac{1}{2^{N_T}} \sum_{s=1}^{2^{N_T}} \int_{\frac{1}{2}  \gamma {P\textbf{h}^T_i \textbf{w}_{i_s} }}^{\infty} \mathcal{N}_{y_i}\left( 0, \sigma^2_i\right) {\rm d}y_i
\end{aligned}
\end{equation}
where $ \textbf{W}_s$ is the precoding matrix  associated with a transmitted symbol $\textbf{x}_s$ and $\mathcal{N}( \mu , \sigma^2)$ denotes the probability density function (PDF) of Gaussian distribution with mean $\mu$ and variance $\sigma^2$.
It is evident that (\ref{equ:BER_CI}) can be expressed  in terms of the one dimensional Gaussian $Q-$function, $\mathcal{Q}(\cdot)$, yielding
\begin{equation}\label{equ:BER_CI2}
\mbox{Pr}_{e_i} = \frac{1}{2^{N_T}} \sum_{s=1}^{2^{N_T}}  \mathcal{Q}\left(\frac{ \gamma P \textbf{h}^T_i \textbf{w}_{i_s} }{2\sigma_i}  \right).
\end{equation}

\subsubsection{CI BER Performance Under Outdated CSI}
Assuming outdated CSI which results from the mobility of users between consecutive CSI updates,  the error probability at the $i^{th}$ PD can be upper bounded  by \eqref{equ:BER_CI_CSI}, at the top of the next page,
\begin{figure*}
\begin{equation}\label{equ:BER_CI_CSI}
 \mbox{Pr}_{e_i}  \leq \frac{1}{2^{N_T}}   \sum_{s=1}^{2^{N_T+1}}    \mathcal{Q}\left( \frac{\gamma  P}{2\sigma_i} \textbf{h}^T_i \textbf{w}_{i_s}  - \frac{\gamma  P}{\sigma_i} \sum_{  k\neq i} \Upsilon_{{ik}_s} A_{sk}\right)   +  \frac{1}{2^{N_T}}   \sum_{s=1}^{2^{N_T+1}}    \mathcal{Q}\left(   \frac{\gamma  P}{2 \sigma_i} \textbf{h}^T_i \textbf{w}_{i_s} +  \frac{\gamma  P}{\sigma_i} \Upsilon_{{ii}_s} +  \frac{\gamma  P}{\sigma_i} \sum_{ k\neq i} \Upsilon_{{ik}_s} A_{sk}\right)
\end{equation}
\hrulefill
\end{figure*}
\normalsize
where $\Upsilon \in \mathbb{R} ^{N_R \times N_R}$ is the residual error matrix, due to the movement of a user between two channel updates, namely
\begin{equation}
 \Upsilon_{{ik}}= h_{i}^T \times \hat{w}_{k}.
 \end{equation}
  Also,  $\hat{w}_{k}$ is the  $k^{th}$ column vector in $\hat{\textbf{W}}$, that is determined based on the  estimate $\hat{\textbf{H}}=\textbf{H} + \mathcal{E}$, with $\mathcal{E}$ denoting the error bound  resulting when a user moves with maximum velocity  between two channel updates.    Finally,
\begin{equation}
\label{equ:A}
\begin{aligned}
\textbf{A}= \left[\begin{array}{ccc}
    A_{1 1}       &  \dots & A_{1  N_T}   \\
    A_{2 1}      &  \dots & A_{2  N_T}   \\
        \vdots        &  \vdots &\vdots    \\
    A_{2^{N_T}  1}       &  \dots & A_{2^{N_T}  N_T} \\
\end{array} \right]
= \left[\begin{array}{cccc}
    0     & 0 & \dots  & 0  \\
    0         & 0 &  \dots  & 1   \\
         \vdots    & \vdots    &\vdots  &\vdots  \\
    1       & 1 &   \dots  & 1 \\
\end{array} \right]
\end{aligned}
\end{equation}
where the elements of $\textbf{A}$ demonstrate the possible combinations of the transmitted OOK symbols.
\subsection{Optical Adaptive Precoding}
It is evident that when the  interference between the different users is constructive, not all  off-diagonal components of $\textbf{H}$ need to be nulled. Thus,   eliminating only the destructive interference  can provide diversity gain for the receivers to avert from the noise enhancement effect. We  refer to this as optical adaptive precoding  (OAP). More specifically, the constructive interference denotes  the ICI that adds to the energy of the symbol of interest, yielding favorable increase in its distance from the constellation thresholds. By also considering  OOK   and recalling  that the channel matrix $\textbf{H}$ in VLC is always positive, it follows that ICI is constructive when the instantaneous symbols are equal.  Thus, using  the knowledge of transmitted data, the interference can be evaluated symbol by symbol and a constructive adaptive matrix  $\textbf{T}$ can be constructed  as follows:
\begin{algorithm}[!h]
\label{algorithm}
\caption{Design of adaptive precoding matrix}
%\ForAll{symbols s=1:$N_{symbols}$}{
\For{{\rm \bf{ all}} symbols s=1:$N_{symbols}$}{
\For{k=1:$N_R$}
{
\For{l=1:$N_R$}
{\If{${x_k}_s$ == ${x_l}_s$} {$T_{kl}$=1}
\Else {$T_{kl}$=0}}}
\textit{Transmit}{$\,\beta\textbf{W} \textbf{T}. x_s$}}
\end{algorithm}

It is noticed that the new adaptive precoding matrix $\textbf{W}_d=\textbf{W}.\textbf{T}$ consists of a  fixed user-level precoding  matrix $\textbf{W}$, which depends on the channels of the individual users, as well as a symbol-level adaptive  precoding matrix $\textbf{T}$ that only depends on the symbols to be transmitted to each user. Furthermore, since  the values of  $\textbf{T}$  are either  zeros or ones, we can use the same scaling factor as in (\ref{equ:scaling}).

It  is evident that the received signal at each PD is a summation of its desired signal and the signals of the LEDs with constructive ICI. Based on this,  the received signal at  $PD_i$, $i=1,2,...,N_R$ is represented as
\begin{equation}
\label{y}
y_{i} = \gamma  P  \sum \limits_{j\in G(i)} \textbf{h}^T_i {\textbf{w}_d}_j x_j + z_{i}
\end{equation}
where $G(i)$ is the group of LEDs from which $PD_i$ receives the desired data and the constructive ICI, which is in fact entirely wasted in  conventional CI. Based on this, it follows that the  corresponding instantaneous SNR can now be expressed as
\begin{equation}
\textrm{SNR}_{i}= \frac{\gamma P}{2\sigma_i}   \sum \limits_{j\in G(i)} \textbf{h}^T_i {\textbf{w}_d}_j x_j.
\end{equation}
Consequently, the normalized overall achievable  throughput of the MIMO VLC system under OAP can be expressed as:
\begin{equation}\label{equ:Th_OAP}
\mbox{Th}_{OAP}=\sum_{i=1}^{N_R} \log_2\left(1+\frac{\gamma P}{2\sigma_i}   \sum \limits_{j\in G(i)} \textbf{h}^T_i {\textbf{w}_d}_j x_j\right).
\end{equation}
\subsubsection{OAP BER Performance Under Perfect CSI}
The error probability at the $i^{th}$ PD can be expressed by \eqref{equ:BER_DCI}, at the top of the next page,
\begin{figure*}
\begin{equation}\label{equ:BER_DCI}
\mbox{Pr}_{e_i}  =  \frac{1}{2^{N_T}} \sum_{s=1}^{2^{N_T}}  \int_{-\infty}^{\frac{1}{2}  \gamma P\textbf{h}^T_i {\textbf{w}_d}_{j_s}  } \mathcal{N}_{y_i}\left( \gamma P \sum \limits_{j\in G(i)} \textbf{h}^T_i {\textbf{w}_d}_{j_s}  + , \sigma^2_i\right) {\rm d}y_i  + \frac{1}{2^{N_T}} \sum_{s=1}^{2^{N_T}}  \int_{\frac{1}{2}  \gamma P\textbf{h}^T_i {\textbf{w}_d}_{j_s}  }^{\infty} \mathcal{N}_{y_i}\left( 0, \sigma^2_i\right) {\rm d}y_i
\end{equation}
\hrulefill
\end{figure*}
which can be expressed in closed-form in terms of the $Q-$function, namely
\begin{equation*}
\mbox{Pr}_{e_i}  = \frac{1}{2^{N_T}} \sum_{s=1}^{2^{N_T}} \mathcal{Q}\left(\frac{ \gamma P  }{2\sigma_i}  \textbf{h}^T_i {\textbf{w}_d}_{i_s}  + \frac{\gamma P }{\sigma_i} \sum \limits_{j\in G(i),j \neq i} \textbf{h}^T_i {\textbf{w}_d}_{j_s}   \right)
\end{equation*}
\begin{equation}  \label{equ:BER_DCI2}
+ \frac{1}{2^{N_T}} \sum_{s=1}^{2^{N_T}} \mathcal{Q}\left(\frac{ \gamma P}{2\sigma_i} \textbf{h}^T_i {\textbf{w}_d}_{i_s}   \right). \hspace{2cm}
\end{equation}

%\begin{figure*}[   \widetilde{}]
%\centering
%\normalsize
%\subfloat[]\centering{\label{fig:ber_spacing}\includegraphics[width=5.7cm,height=5.9cm]{BER_d.eps}}
%\subfloat[]\centering{\includegraphics[width=5.7cm,height=5.9cm]{theta.eps}
%\label{fig:ber_theta}}
%\subfloat[]\centering{\includegraphics[width=5.7cm,height=5.9cm]{outdated.eps}
%\label{fig:outdated}}
%\caption{BER performance under  (a) different LEDs' spacings, (b) varying  transmitting angle and (c) Outdated CSI.}
%\label{fig}
%\end{figure*}
\subsubsection{OAP  BER Performance Under Outdated CSI}
By following the same methodology as in the case of CI, it follows that   the  BER of OAP under    outdated CSI, can be upper bounded by  (\ref{equ:BER_DCI_CSI}), at the top of the next page.
\begin{figure*}
\begin{equation}\label{equ:BER_DCI_CSI}
\hspace{-0.1cm} \mbox{Pr}_{e_i} \leq \frac{1}{2^{N_T}}    \sum_{s=1}^{2^{N_T+1}}    \mathcal{Q}\left(  \frac{\gamma  P}{\sigma_i}  \left(\frac{1}{2} \textbf{h}^T_i \textbf{w}_{d_{i_s}}  -   \sum_{k\neq i} \Upsilon_{{ik}_s} A_{sk}\right)\right)
 + \frac{1}{2^{N_T}}  \mathcal{Q}\left(   \frac{\gamma  P}{\sigma_i} \left( \frac{1}{2} \textbf{h}^T_i \textbf{w}_{d_{i_s}}  +  \sum \limits_{j\in G(i)} \textbf{h}^T_i {\textbf{w}_d}_{j_s} +  \Upsilon_{{ii}_s} + \sum_{k\neq i} \Upsilon_{{ik}_s} A_{sk}\right) \right)
\end{equation}
\hrulefill
\end{figure*}
%Under perfect CSI, all the destructive interference will be completely eliminated, and the constructive interference will result in better detection compared to conventional CI.
\section{Numerical Results}
\label{sec:results}
\begin{figure*}[   \widetilde{}]
\normalsize
\subfloat[]{\label{ref_label1}\includegraphics[width=2.5in,height=2.4in]{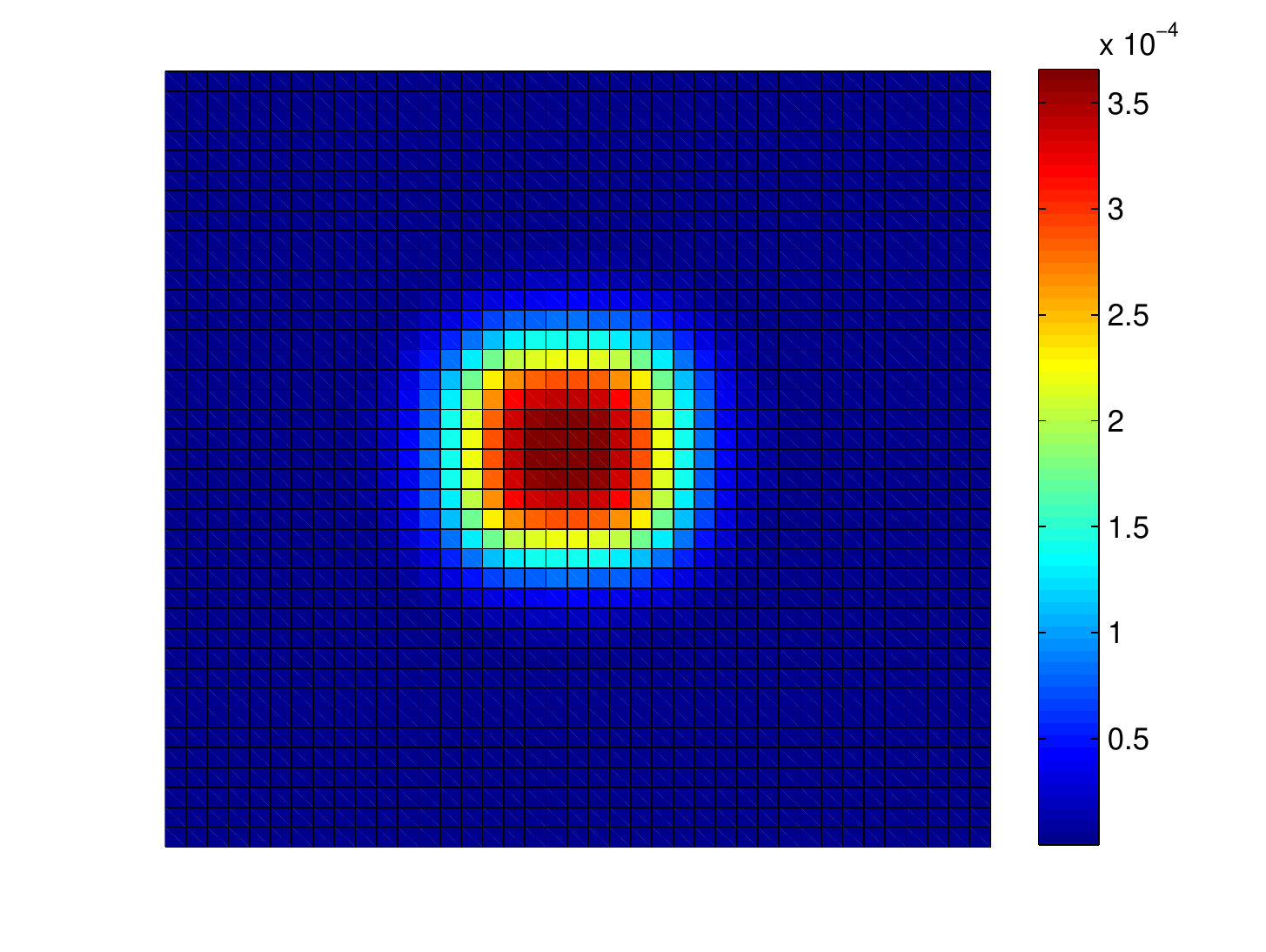}}
  \subfloat[]{\label{ref_label2}\includegraphics[width=2.5in,height=2.4in]{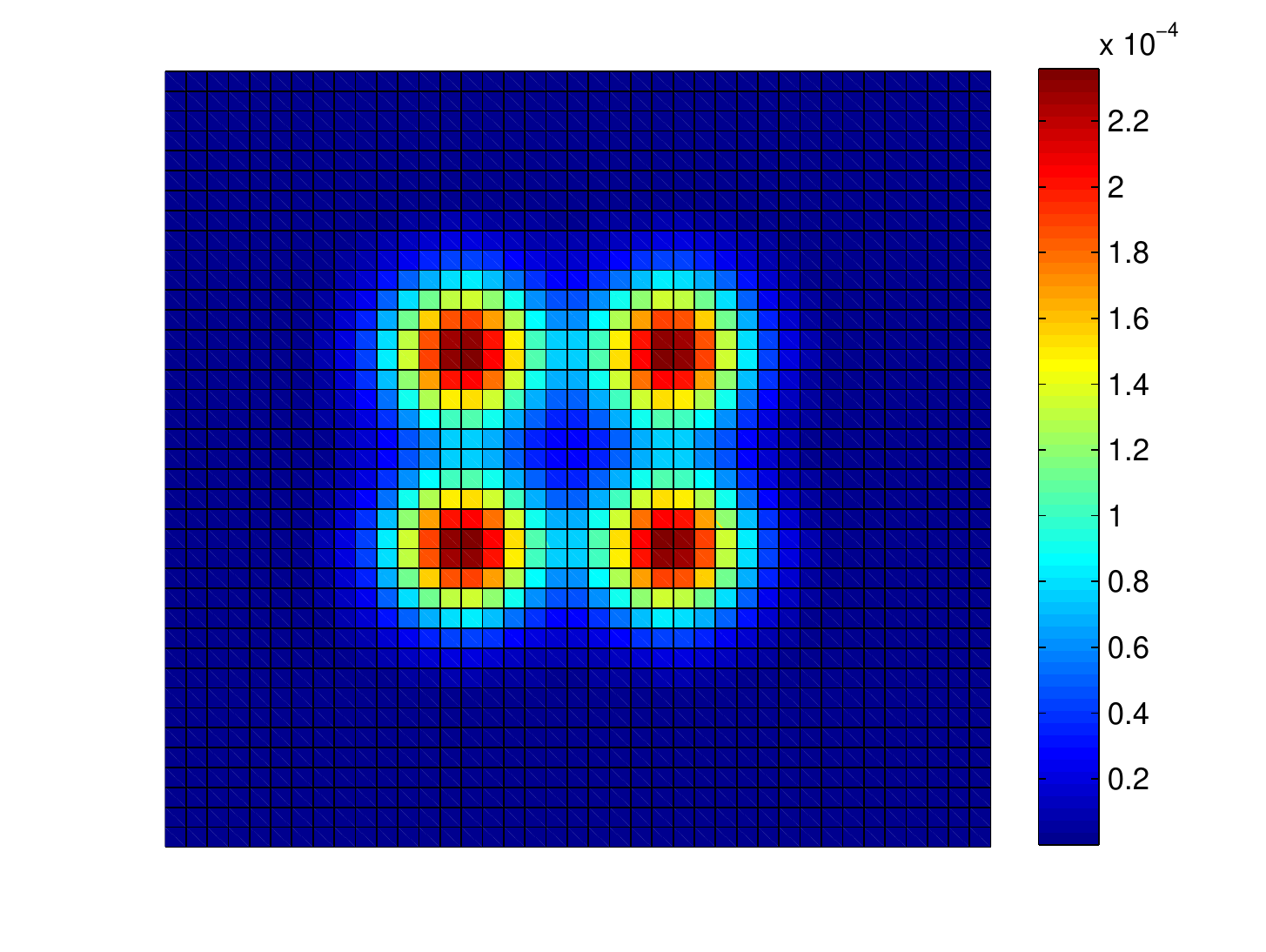}}
  \subfloat[]{\label{ref_label2}\includegraphics[width=2.5in,height=2.4in]{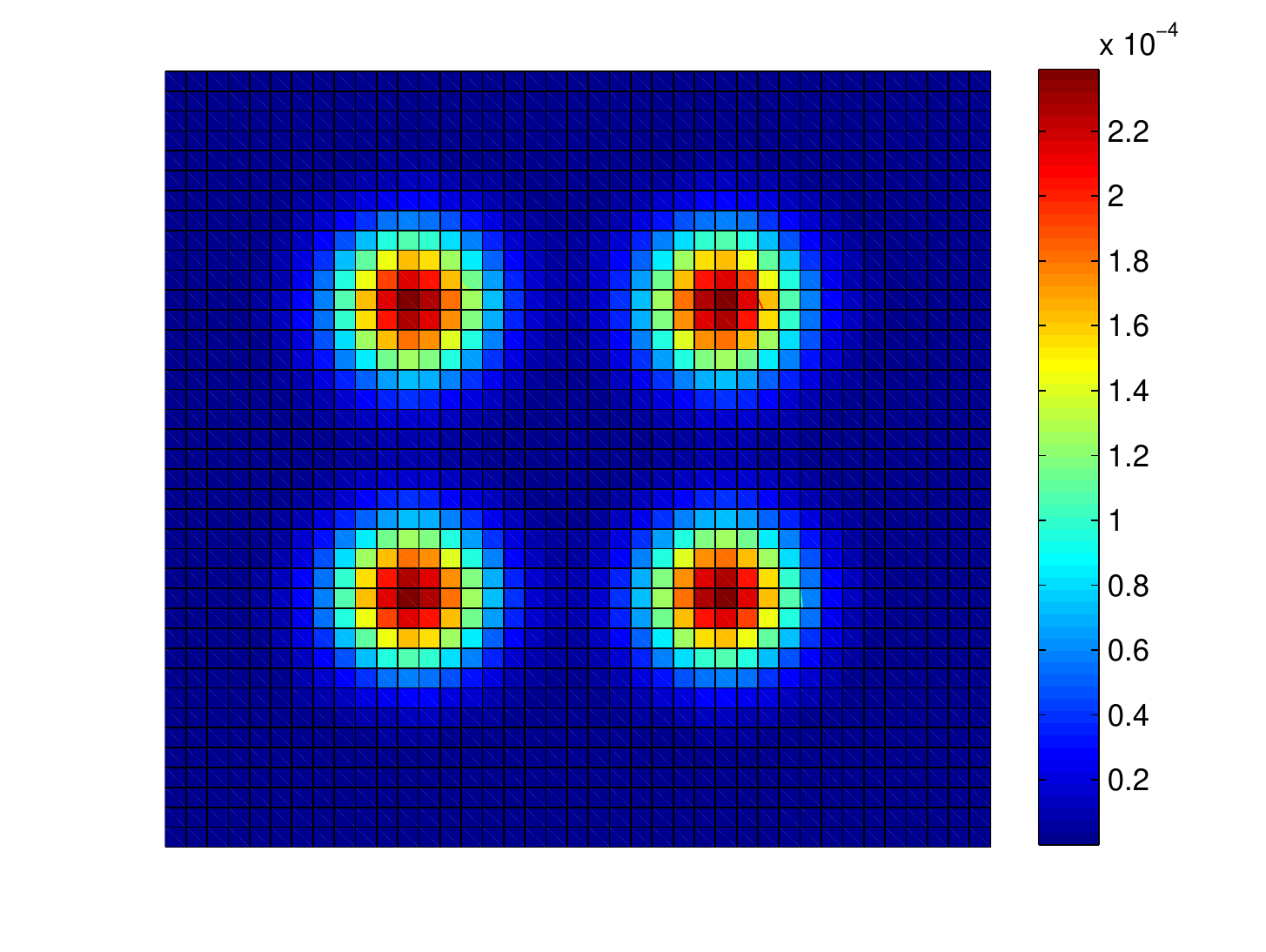}}
  \caption{\label{fig:h}VLC channel gains  for four transmitting LEDs with $\varphi_{1/2}=15^\circ$ and spacing of (a) 0.5 m, (b) 1.0 m and (c) 2.0 m.}
\noindent\makebox[\linewidth]{\rule{520pt}{.4pt}}
\end{figure*}

In this section, we analyze the corresponding  results for the  conventional CI and the proposed adaptive  precoding. This is realized by both analytic (solid lines) and respective computer simulation (markers) results.
To this end, we consider  a $4.0m \times 4.0m \times 3.0m$ indoor room scenario and  an indicative  $4 \times 4$ MIMO setup with different LEDs' spacings on the x- and y-axis of $0.25$m, $0.5$m and $1.0$m\footnote{It is noted here  that the considered specifications are indicative of a practical application in realistic indoor environments, while the proposed results are generic and not subject to any physical dimensional restrictions.} . The corresponding channel gains across the room for the three different setups are  shown in Fig. \ref{fig:h} for a receiver plane $z=0.75$ m.   Also, the  locations of the users are assumed to be fixed and aligned with their respective transmitters and unless otherwise specified, the system parameters in  Table \ref{Tab:Parameters} are used in the computer simulations.

 \begin{table}[t]
\caption{SIMULATION PARAMETERS}
\centering
\renewcommand{\arraystretch}{1.3}
\setlength{\tabcolsep}{20pt}
\begin{tabular}{l r}
\multicolumn{2}{c}{Transmitters Parameters} \\
\hline
Number of LEDs per luminary&$ 60 \times 60$ \\[2ex]\hline
Transmitted  power per  LED & $10$ mW  \\[1.25ex]\hline
Transmitter semi-angle $\varphi_{1/2}$&$ 15^\circ$  \\[1.25ex]\hline\\
\multicolumn{2}{c}{Receivers Parameters} \\
\hline
Receiver FOV &$15^\circ$  \\[1.25ex]\hline
Physical area of PD& $1.0$ $cm^{2}$ \\[1.25ex]\hline
PD responsivity $\gamma$&$ 1$ A/W \\[1.25ex]\hline
Refractive index of PD Lens&$ 1.5$ \\[1.25ex]\hline
Gain of optical filter $Ts(\phi)$&$ 1.0$ \\[1.25ex]\hline
Background current $I_{bg}$ &$100 \mu A $\\[1.25ex]\hline
Noise bandwidth factor $I_2$& $0.562$\\[1.25ex]
\hline
\end{tabular}
\label{Tab:Parameters}
\end{table}

\begin{figure}[t]
\centering
\includegraphics[width=9cm,height=7cm]{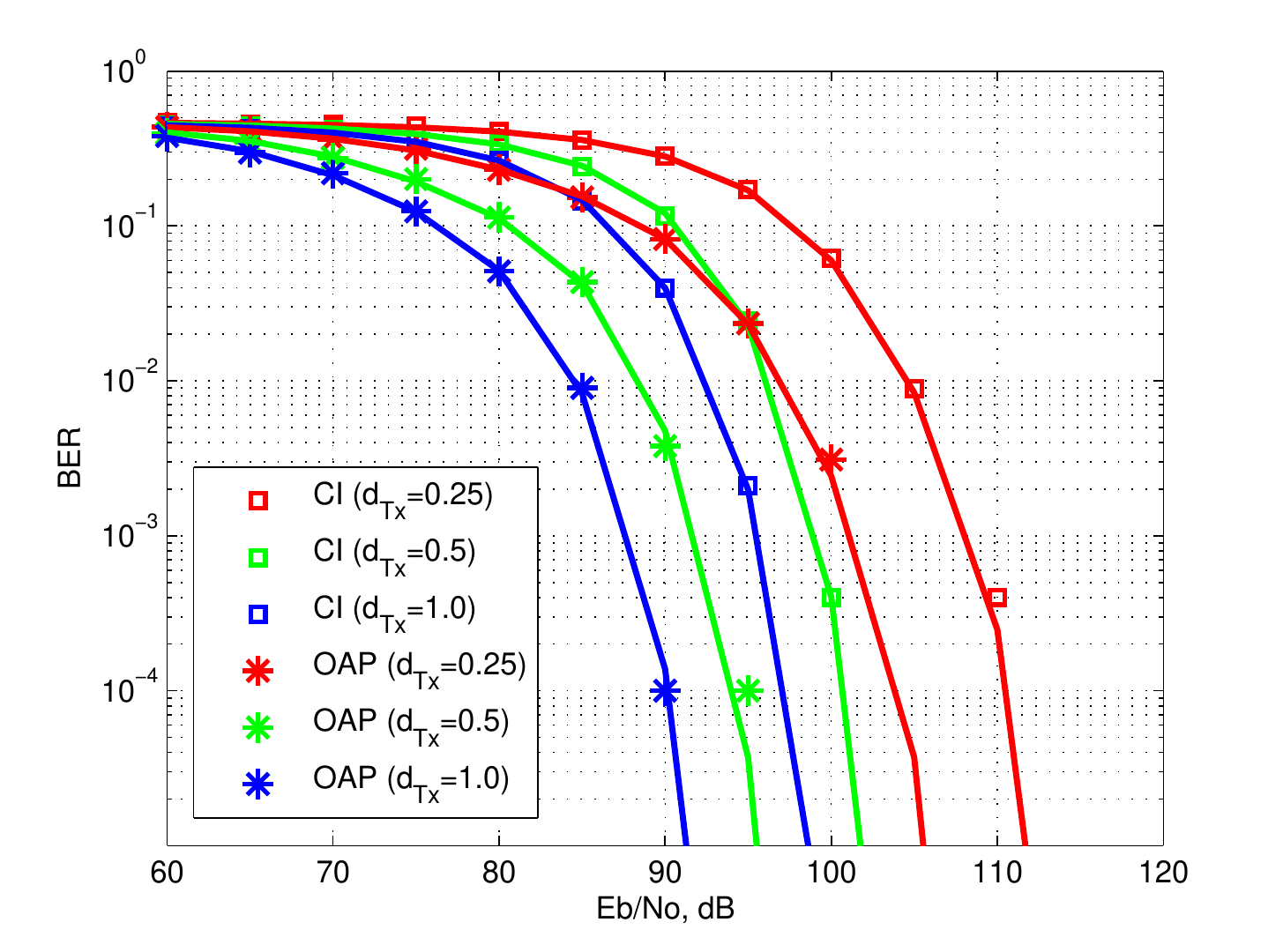}
\caption{Effect of channel correlation on the BER performance  of CI and OAP.}
\label{fig:ber_spacing}
\end{figure}

 \label{sec:results}
%The locations of the users are assumed to be fixed and aligned with their respective transmitters.
%
 Fig. \ref{fig:ber_spacing} illustrates  the BER of CI and OAP for the three considered scenarios. It is shown that the proposed OAP outperforms the conventional CI regardless of the transmitters spacings. In fact, it is noticed that  CI requires an SNR increase of    about $8$ dB  to achieve the   BER of the  OAP counterpart. Furthermore, the highest error rate occurs  at the smallest  spacing, which is due to the highly correlated links that increase the  severity of the noise enhancement induced at the receivers. %On the contrary, OAP provides enhanced performance in the case of small transmitters spacing. This is due to the fact  that OAP benefits from the constructive interference, and hence provides lower BER in the case of high channel correlation.
It is also shown that the analytical and simulation results are in tight agreement, which verifies the validity of the derived expressions.

\begin{figure}[t]
\centering
\includegraphics[width=9.0cm,height=7cm]{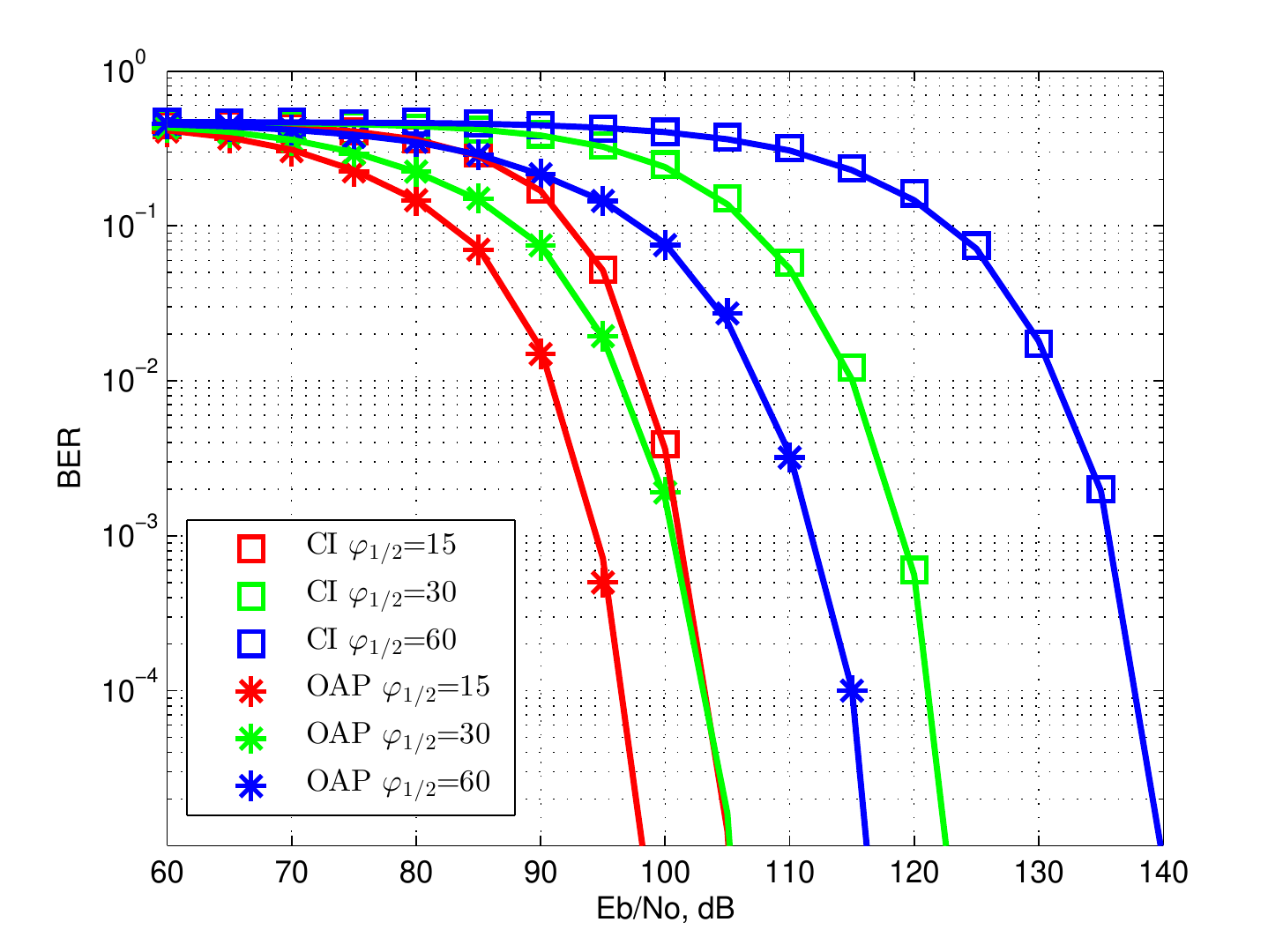}
\caption{Effect of varying the LED transmitting angle $\varphi_{1/2}$ on the BER performance of CI and OAP.}
\label{fig:ber_theta}
\end{figure}

Next, we investigate the effect of changing  the transmitters semi-angle at half power, $\varphi_{1/2}$, where fixed transmitters spacing of $1.0$m is assumed.    Larger $\varphi_{1/2}$  alleviates the  transmitted beam concentration. In the case of conventional CI, this has a detrimental  effect on the system performance for two reasons:  i) it reduces the channel gains of the users leading to lower received power; and ii)  it increases the ICI that needs to be cancelled, which practically creates more noise enhancement at the receivers.
Fig. \ref{fig:ber_theta}  demonstrates the performance degradation of CI when larger transmitting angles are used, which indicates   that the effect of increasing $\varphi_{1/2}$ is less significant in the case of adaptive precoding. This is mainly because the effect of the reduction of the channel gains of the users is offset by the constructive interference that adds to the SNR at the receivers. \\

\begin{figure}[t]
\centering
\includegraphics[width=9.0cm, height=7cm]{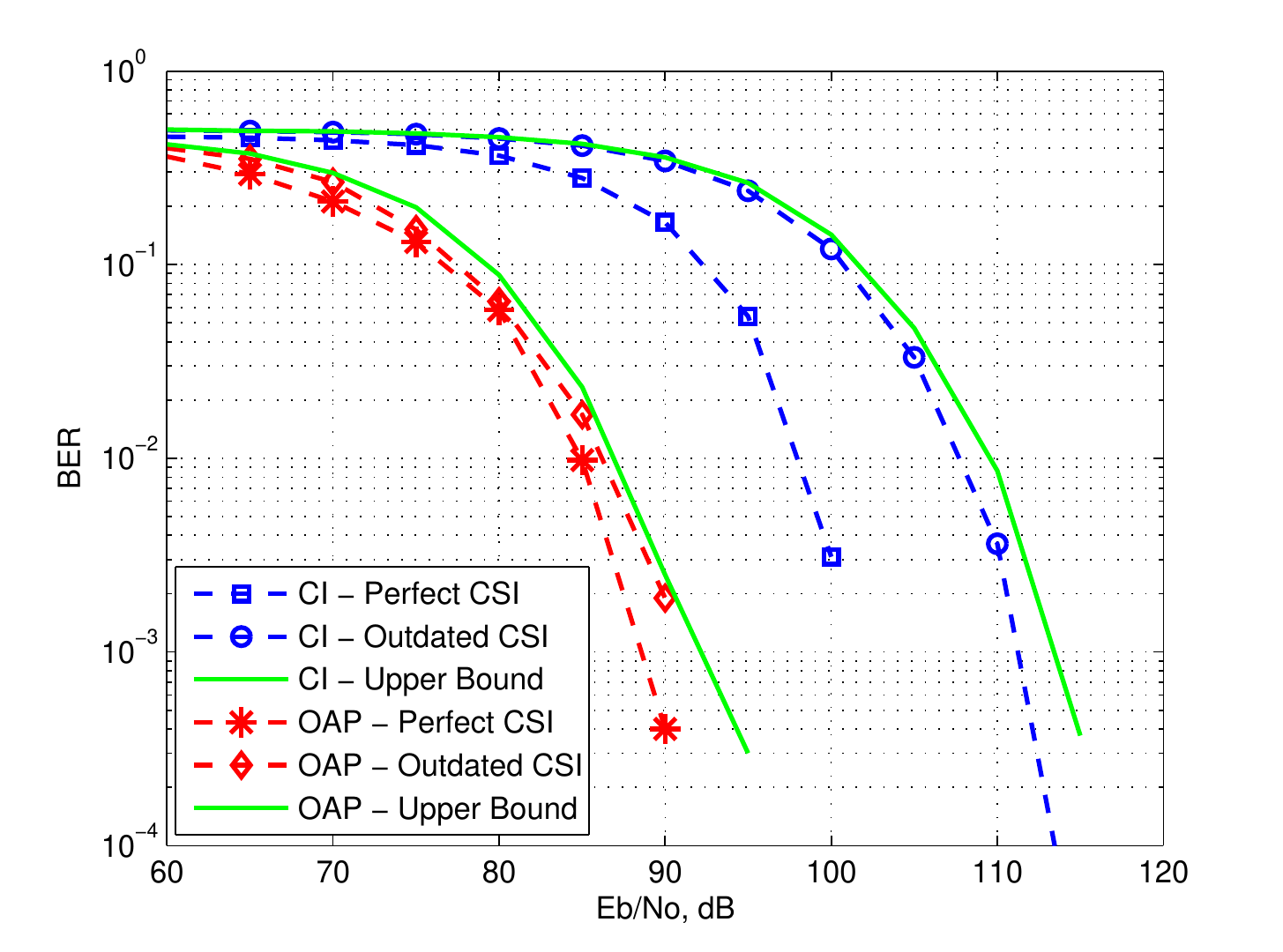}
\caption{BER vs transmit SNR with outdated CSI.}
\label{fig:outdated}
\end{figure}

Next, Fig. \ref{fig:outdated} illustrates the effect of outdated CSI on the system performance, where the proposed  error bound is computed assuming  maximum velocity of users between two channel updates. It is evident that adaptive precoding is more robust to CSI error, especially  in the low SNR region. On the contrary, the  effect of imperfect CSI becomes more significant at high SNR as the noise is no longer dominating, and thus channel uncertainty becomes a limiting factor.
\indent

\begin{figure}[t]
\centering
\includegraphics[width=9.0cm, height=7cm]{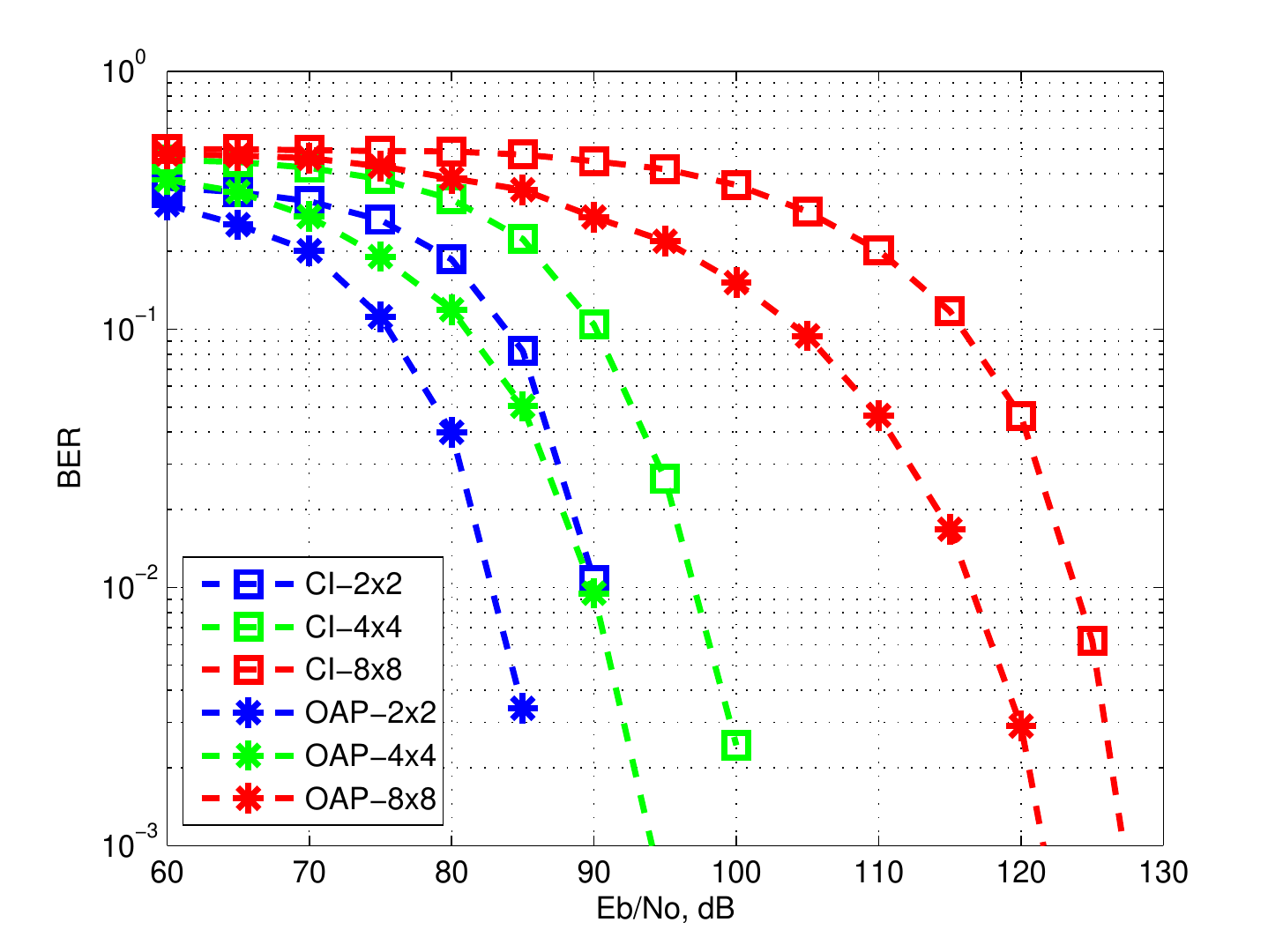}
\caption{BER vs transmit SNR for different MIMO configrations.}
\label{fig:mimo}
\end{figure}
\begin{figure}[t]
\centering
\includegraphics[width=9.0cm, height=7cm]{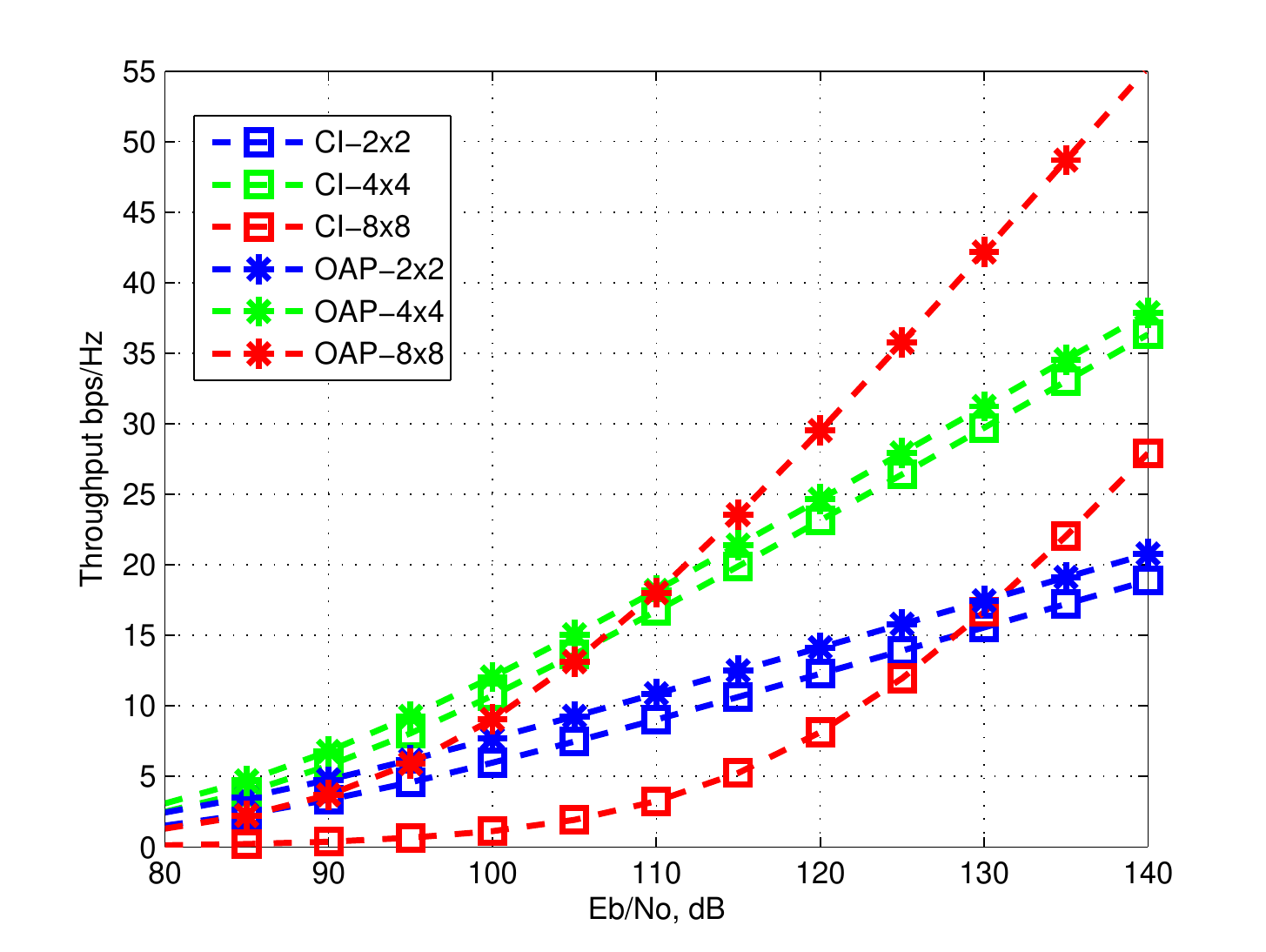}
\caption{Normalized Throughput vs transmit SNR for different MIMO configrations.}
\label{fig:th}
\end{figure}

\indent
In the following, we examine the performance of the VLC MIMO system under different number of transmitting LEDs. To this end,  Fig. \ref{fig:mimo} shows the BER performance for $2\times2$, $4\times4$ and $8\times8$ MIMO configurations. As expected, the BER performance degrades as the number of transmitters increases, which is due to the severity of the noise enhancement caused by the high correlation between the spatial subchannels. It is evident that OAP provides noticeable performance enhancement compared to conventional CI. Moreover, the normalized achievable system throughput for the three MIMO configurations is shown in Fig. \ref{fig:th}. As expected,  increasing the number of transmitters from $2$ to $4$ leads to increased achievable  throughput for both schemes. It is noted, however, that the system throughput under conventional CI degrades significantly for the case of $8$ transmitters  despite the increase in the number of users. This is due to the noise enhancement that   compromises the achievable system throughput. On the contrary, the proposed OAP provides substantial throughput enhancement for the $8\times8$ MIMO setup.
\indent

\section{Conclusions}
\label{sec:conc}
%In this paper, we demonstrated that the performance of indoor MU-MIMO VLC systems can be significantly enhanced by exploiting the constructive part of the interference exiting between the spatial optical sub-channels.%
In this paper, the knowledge of   transmitted data was exploited to  design a low-complexity adaptive precoder based on channel inversion. The presented  analysis and respective simulations showed that  the proposed precoding scheme provides  considerable  performance enhancement compared to conventional channel inversion.  In this context,  it was shown that the concept of adaptive precoding creates new opportunities in VLC systems, as the classification of interference into constructive and destructive is independent of CSI due to the nature of the optical channel. Future work will attempt extensions to higher-order modulations, which is a non-trivial task as it requires a   thorough design of the precoding matrix.
\label{sec:conc}
\bibliographystyle{IEEEtran}
\bibliography{referencevlc}

\begin{thebibliography}{10}
\providecommand{\url}[1]{#1}
\csname url@samestyle\endcsname
\providecommand{\newblock}{\relax}
\providecommand{\bibinfo}[2]{#2}
\providecommand{\BIBentrySTDinterwordspacing}{\spaceskip=0pt\relax}
\providecommand{\BIBentryALTinterwordstretchfactor}{4}
\providecommand{\BIBentryALTinterwordspacing}{\spaceskip=\fontdimen2\font plus
\BIBentryALTinterwordstretchfactor\fontdimen3\font minus
  \fontdimen4\font\relax}
\providecommand{\BIBforeignlanguage}[2]{{%
\expandafter\ifx\csname l@#1\endcsname\relax
\typeout{** WARNING: IEEEtran.bst: No hyphenation pattern has been}%
\typeout{** loaded for the language `#1'. Using the pattern for}%
\typeout{** the default language instead.}%
\else
\language=\csname l@#1\endcsname
\fi
#2}}
\providecommand{\BIBdecl}{\relax}
\BIBdecl

\bibitem{cisco_indoor}
``Cisco service provider wi-fi: A platform for business innovation and revenue
  generation, {Cisco, San Jose, CA, USA}, 2012.''
  \url{http://www.cisco.com/en/US/solutions/collateral/ns341/ns524/ns673/solution-overview-c22-642482.html/},
  accessed: 2016-01-16.

\bibitem{VLC_mag}
H.~Burchardt, N.~Serafimovski, D.~Tsonev, S.~Videv, and H.~Haas, ``{VLC}:
  Beyond point-to-point communication,'' \emph{IEEE Commun. Mag.}, vol.~52,
  no.~7, pp. 98--105, July 2014.

\bibitem{market}
A.~Jovicic, J.~Li, and T.~Richardson, ``Visible light communication:
  opportunities, challenges and the path to market,'' \emph{IEEE Commun. Mag.},
  vol.~51, no.~12, pp. 26--32, Dec. 2013.

\bibitem{survey2}
P.~H. Pathak, X.~Feng, P.~Hu, and P.~Mohapatra, ``Visible light communication,
  networking, and sensing: A survey, potential and challenges,'' \emph{IEEE
  Commun. Surveys Tuts.}, vol.~17, no.~4, pp. 2047--2077, Fourthquarter 2015.

\bibitem{haas_book}
S.~Dimitrov and H.~Haas, \emph{Principles of LED Light Communications: Towards
  Networked Li-Fi}.\hskip 1em plus 0.5em minus 0.4em\relax Cambridge University
  Press, 2015.

\bibitem{murat2}
F.~Miramirkhani and M.~Uysal, ``Channel modeling and characterization for
  visible light communications,'' \emph{IEEE Photon. J.}, vol.~7, no.~6, pp.
  1--16, Dec. 2015.

\bibitem{state-of-the-art}
D.~Karunatilaka, F.~Zafar, V.~Kalavally, and R.~Parthiban, ``{LED} based indoor
  visible light communications: State of the art,'' \emph{IEEE Communications
  Surveys Tutorials}, vol.~17, no.~3, pp. 1649--1678, thirdquarter 2015.

\bibitem{MIMO1}
L.~Zeng, D.~O'brien, H.~Minh, G.~Faulkner, K.~Lee, D.~Jung, Y.~Oh, and E.~T.
  Won, ``High data rate multiple input multiple output {(MIMO)} optical
  wireless communications using white {LED} lighting,'' \emph{{IEEE} J. Sel.
  Areas Commun.}, vol.~27, no.~9, pp. 1654--1662, Dec. 2009.

\bibitem{MIMO2}
S.~Navidpour, M.~Uysal, and M.~Kavehrad, ``{BER} performance of free-space
  optical transmission with spatial diversity,'' \emph{{IEEE} Trans. Wireless
  Commun.}, vol.~6, no.~8, pp. 2813--2819, Aug. 2007.

\bibitem{MIMO5}
W.~Popoola, E.~Poves, and H.~Haas, ``Spatial pulse position modulation for
  optical communications,'' \emph{J. Lightw. Technol.}, vol.~30, no.~18, pp.
  2948--2954, Sep. 2012.

\bibitem{MIMO7}
Y.-J. Zhu, W.-F. Liang, J.-K. Zhang, and Y.-Y. Zhang, ``Space-collaborative
  constellation designs for {MIMO} indoor visible light communications,''
  \emph{{IEEE Photon. Technol. Lett.}}, vol.~27, no.~15, pp. 1667--1670, Aug.
  2015.

\bibitem{MIMO8}
A.~Nuwanpriya, S.~Ho, and C.~Chen, ``Indoor {MIMO} visible light
  communications: Novel angle diversity receivers for mobile users,''
  \emph{IEEE J. Sel. Areas Commun.}, vol.~33, no.~9, pp. 1780--1792, Sep. 2015.

\bibitem{MIMO_VLC_Haas}
T.~Fath and H.~Haas, ``Performance comparison of {MIMO} techniques for optical
  wireless communications in indoor environments,'' \emph{IEEE Trans. Commun.},
  vol.~61, no.~2, pp. 733--742, Feb. 2013.

\bibitem{marshoud}
H.~Marshoud, D.~Dawoud, V.~Kapinas, G.~Karagiannidis, S.~Muhaidat, and
  B.~Sharif, ``{MU-MIMO} precoding for {VLC} with imperfect {CSI},'' in
  \emph{Proc. IEEE 4th International Workshop on Optical Wireless
  Communications {(IWOW)}}, Sep. 2015, pp. 93--97.

\bibitem{BD_VLC}
Y.~Hong, J.~Chen, Z.~Wang, and C.~Yu, ``Performance of a precoding {MIMO}
  system for decentralized multiuser indoor visible light communications,''
  \emph{{IEEE} Photon. J.}, vol.~5, no.~4, pp. 7\,800\,211--7\,800\,211, Aug.
  2013.

\bibitem{THP}
J.~Chen, N.~Ma, Y.~Hong, and C.~Yu, ``On the performance of {MU-MIMO} indoor
  visible light communication system based on {THP} algorithm,'' in \emph{Proc.
  IEEE International Conference on Communications in China {(ICCC)}}, Oct.
  2014, pp. 136--140.

\bibitem{ook_standard}
``{IEEE }standard for local and metropolitan area networks--part 15.7:
  Short-range wireless optical communication using visible light,'' \emph{in
  IEEE Std 802.15.7-2011}, pp. 1--309, Sep. 2011.

\bibitem{Dynamic1}
C.~Masouros and E.~Alsusa, ``Dynamic linear precoding for the exploitation of
  known interference in {MIMO} broadcast systems,'' \emph{{IEEE} Trans.
  Wireless Commun.}, vol.~8, no.~3, pp. 1396--1404, Mar. 2009.

\bibitem{Dynamic2}
C.~Masouros, ``Correlation rotation linear precoding for {MIMO} broadcast
  communications,'' \emph{{IEEE} Trans. Signal Process}, vol.~59, no.~1, pp.
  252--262, Jan. 2011.

\bibitem{coordinated}
H.~Ma, L.~Lampe, and S.~Hranilovic, ``Coordinated broadcasting for multiuser
  indoor visible light communication systems,'' \emph{{IEEE} Trans. Commun.},
  vol.~63, no.~9, pp. 3313--3324, Sep. 2015.

\bibitem{fundamental}
T.~Komine and M.~Nakagawa, ``Fundamental analysis for visible-light
  communication system using {LED} lights,'' \emph{IEEE Trans. Consum.
  Electron.}, vol.~50, no.~1, pp. 100--107, Feb. 2004.

\end{thebibliography}
\balance
\end{document}